\documentclass{PoS}

\usepackage{subcaption}
\usepackage{graphicx}
\usepackage{placeins}
\usepackage{amsmath}

\newcommand{\gevc}{\mathrm{GeV}/c}
\newcommand{\gev}{\mathrm{GeV}}

\newcommand{\mev}{\mathrm{MeV}}

\newcommand {\pT}        {\ensuremath{p_{\mathrm{T}}}}

\newcommand{\pt}{\pT}
\newcommand {\sqrtSnn}   {\ensuremath{\sqrt{s_{\textsc{nn}}}}}
\newcommand {\sqrtS}     {\ensuremath{\sqrt{s}}}

\newcommand {\pp}        {\mbox{$\mathrm {p\kern-0.05em p}$}}
\newcommand {\ppBoldMath} {\mbox{$\mathrm { \mathbf p\kern-0.05em \mathbf p }$}}

\newcommand {\PbPb}      {\ensuremath{\mbox{Pb--Pb} }}

\newcommand {\pPb}       {\ensuremath{\mbox{p--Pb}}}

\newcommand {\tev}      {\mbox{${\rm TeV}$}}

\newcommand {\barn}     {\, \mbox{${\rm barn}$}}

\newcommand {\dg}       {\mbox{$\kern+0.1em ^\circ$}}

\newcommand {\ptch}{\ensuremath{p}_{\mathrm{T,jet}}^{\mathrm{ch}}}

\newcommand {\Ntrig} {N_{\mathrm{trig}}}

\newcommand {\Njet} {N_{\mathrm{jet}}}
\newcommand {\etajet} {\eta_{\mathrm{jet}}}

\newcommand {\diff}{\mathrm{d}}

\title{Current and future measurements of semi-inclusive hadron+jet distributions with ALICE}

\ShortTitle{hadron+jet measurements with ALICE}

\author{\speaker{Jaime Norman}, for the ALICE collaboration
	\\
        Laboratoire de Physique Subatomique et Cosmologie, Grenoble, France\\
        E-mail: \email{jaime.norman@cern.ch}}

\abstract{

The measurement of jets recoiling from a trigger hadron in heavy-ion collisions can be used to understand the properties of the Quark Gluon Plasma. Jet-medium interactions cause jets to lose energy in the medium and may modify the jet structure. Jet deflection towards large angles due to scattering off of quasi-particles in the Quark-Gluon Plasma may also occur, which can be studied through a measurement of the hadron-jet acoplanarity. These phenomena can be studied through the semi-inclusive distribution of track-based jets recoiling from a trigger hadron.
This contribution presents the latest measurements and prospects of semi-inclusive hadron+jet distributions with ALICE. Constraints on energy loss in $\pPb$ collisions and future prospects to measure energy loss in smaller systems are shown. A jet shape measurement of N-subjettiness using recoil jets is outlined. Finally, prospects for hadron+jet acoplanarity measurements with ALICE are presented.

}
\FullConference{International Conference on Hard and Electromagnetic Probes of High-Energy Nuclear Collisions\\
		30 September - 5 October 2018\\
		Aix-Les-Bains, Savoie, France}

\begin{document}

\section{Introduction}
\label{sec:intro}

Measurements of jets created in ultrarelativistic heavy-ion collisions provide unique probes to characterise the hot and dense QCD medium created in these collisions. The measurement of inclusive jets (see e.g. \cite{full-jets-these} for recent ALICE results) show a significant suppression in heavy-ion collisions with respect to pp collisions, indicating that partons lose energy while travelling through and interacting with the QCD medium.

The measurement of jets recoiling from a trigger hadron is being employed to further study jet quenching effects.
ALICE has measured the trigger-normalised semi-inclusive yield of jets recoiling from a trigger hadron $\frac{1}{\Ntrig}\frac{\diff^2\Njet}{\diff\ptch \diff \etajet}$, differential in jet transverse momentum $\ptch$. A variable $\Delta_\mathrm{recoil} $ is then defined as the difference between the trigger-normalised recoil jet distributions in `reference' and `signal' trigger track $\pt$ intervals $TT_\mathrm{ref}$ and $TT_\mathrm{sig}$ \cite{Adam:2015doa}:


\begin{equation}
\Delta_\mathrm{recoil} = \frac{1}{\Ntrig}\frac{\diff^2\Njet}{\diff\ptch \diff \etajet} \Bigg\rvert_{TT_\mathrm{sig}} -  c_\mathrm{ref} \cdot
\frac{1}{\Ntrig}\frac{\diff^2\Njet}{\diff\ptch  \diff \etajet} \Bigg\rvert_{TT_\mathrm{ref}} 
\label{eq:DeltaRecoil}
\end{equation}


\noindent where $c_\mathrm{ref}$ accounts for the combined effects of invariance of total jet yield with trigger track $\pt$. With this observable one removes entirely the background from uncorrelated reconstructed jets, giving the possibility of extending jet measurements to low-$\pt$ and high jet resolution parameter $R$.
The jet population measured with this technique is not biased in terms of jet fragmentation pattern.
As a trigger-normalised and semi-inclusive quantity one can also avoid model-dependent assumptions to relate event activity to event geometry, leading to greater systematic sensitivity to jet quenching effects in small systems \cite{Acharya:2017okq}.
It is noted that the measurements shown here use jets reconstructed from charged tracks only, i.e. `track-based jets'.

\section{Constraints on jet quenching in smaller systems}
\label{sec:quenching}

\begin{figure}[hbt]
	\centering
	\includegraphics[width=.4\textwidth]{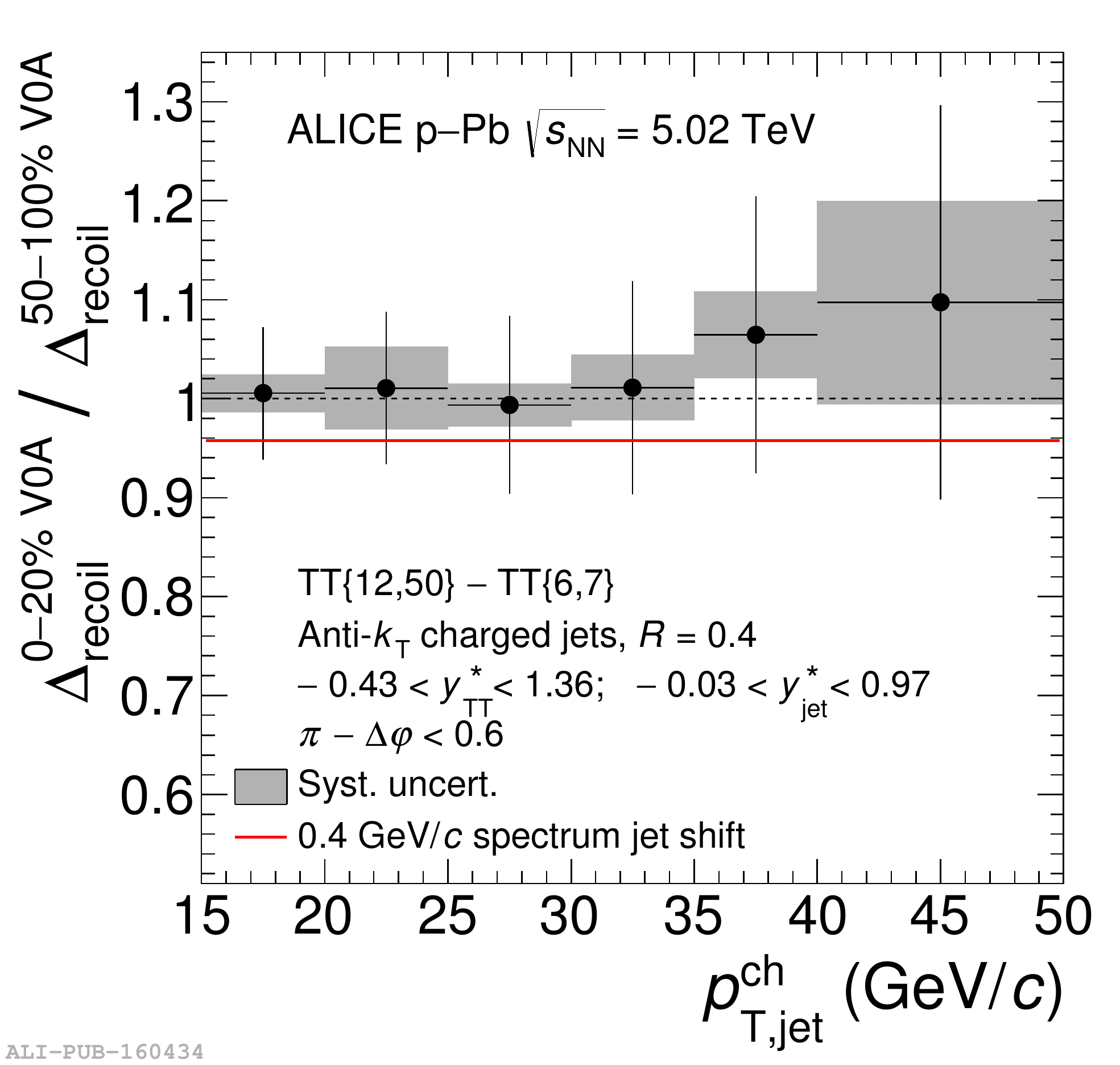}
	\includegraphics[width=.4\textwidth]{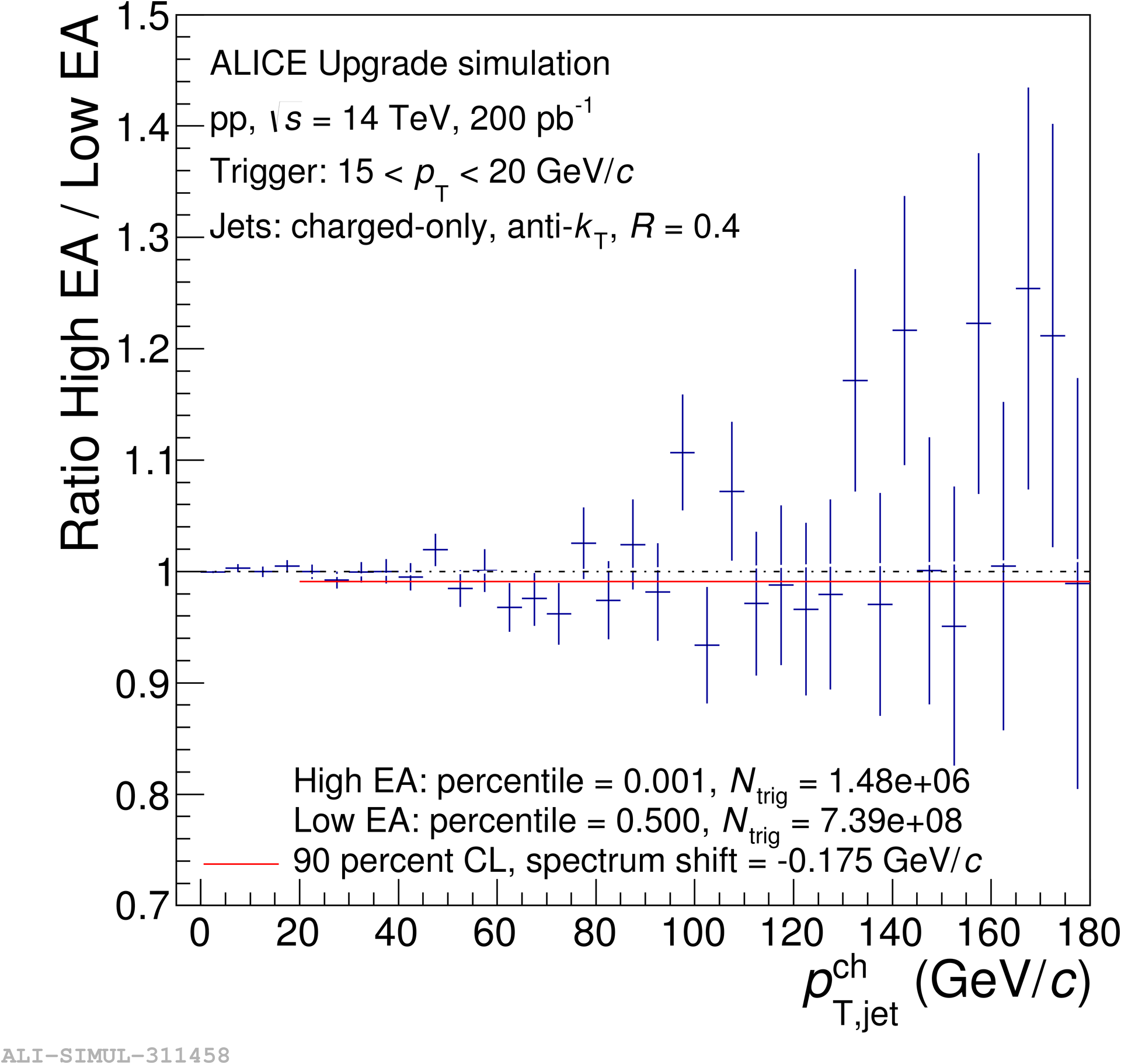}
	\caption[]{Left: The ratio of $\Delta_\mathrm{recoil}$ in high/low event activity classes in $\pPb$ collision at $\sqrtSnn = 5.02 ~\tev$. Right: Projection of the ratio of $\Delta_\mathrm{recoil}$ in high/low event activity classes in pp collisions at $\sqrtS = 14 ~\tev$ in Run 3/4 of the LHC. The red line in both figures corresponds to the $90\%$ CL spectrum shift. }  
	\label{fig:small-system-quenching}
\end{figure}

The measurement of the trigger-normalised recoil jet distributions in $\PbPb$ collisions indicates that jets lose a significant amount of energy in $\PbPb$ collisions, and that this energy is predominantly radiated to angles greater than $R=0.5$ \cite{Adam:2015doa}. A similar analysis was performed in $\pPb$ collisions in different event-activity classes, defined using the signal magintude in the V0A detectors of ALICE, to test whether jets are quenched in smaller systems \cite{Acharya:2017okq}. Figure \ref{fig:small-system-quenching} (left) shows the $\Delta_\mathrm{recoil}$ ratio in high/low event activity classes for recoil jets with $R=0.4$ from $15-50~\gevc$. The ratio is consistent with unity, indicating minimal jet quenching in $\pPb$ collisions, and a limit of $<0.4 ~\gev$ out-of-cone energy loss for jets in this $\pt$ range is set (90\% CL).

The sensitivity to jet quenching in small systems in Run 3/4 of the LHC has also been assessed, based on PYTHIA simulations to estimate the expected number of charged hadron triggers and trigger normalised recoil jet spectrum for a given integrated luminosity. Figure \ref{fig:small-system-quenching} (right) shows the projection of the same observable for $R=0.4$ jets in pp collisions at $\sqrtS = 14 ~\tev$ with an integrated luminosity of 200 p$\barn^{-1}$, where the ratio of `central' and `peripheral' events corresponds to the 0--0.1\% centrality percentile and 50--100\% centrality percentile respectively. Here no event-activity shift is included, and the statistical limit (90\% CL) on a measurement of out-of-cone energy loss is 175 $\mev$ for this system. For $\pPb$ collisions where the high-EA is set to the 0--5\% percentile, this limit is 70 $\mev$. While the corresponding systematic uncertainties are not estimated, it is noted that the statistical uncertainties were dominant in the Run 1 measurement.

\section{Substructure of recoil jets}
\label{sec:substructure}

The measurement of jet shapes and their modification in heavy-ion collisions can give  further insight into jet quenching mechanisms. For example, the role of colour coherence \cite{Mehtar-Tani:2016aco} can be probed by studying how 2-pronged jets are modified in heavy-ion collisions. The N-prongness of jets is measured through the N-subjettiness observable $\tau_N$. For this observable, jets are reclustered using an exclusive clustering algorithm, and `subjet' axes are found by unwinding the last clustering step. $\tau_N$ is then defined as

\begin{equation}
\tau_N = \frac{\sum_{i=1}^N p_{\mathrm{T,i}} \mathrm{min}(\Delta R_{\mathrm{i,1}} , \Delta R_{\mathrm{i,2}},..., \Delta R_{\mathrm{i,N}})}{R_0 \sum_{i=1}^N (p_{\mathrm{T,i}})}
\end{equation}

\noindent where $\Delta R_{\mathrm{i,j}}$ is the $\phi - \eta$ distance between track $i$ and subjet $j$, $p_{\mathrm{T,i}}$ is the $\pt$ of the $i$-th jet constituent and $R_0$ is the jet resolution parameter. The ratio $\tau_2 / \tau_1$ is a measure of how 2-pronged a jet is. 

\begin{figure}[hbt]
	\centering
	\includegraphics[width=.46\textwidth]{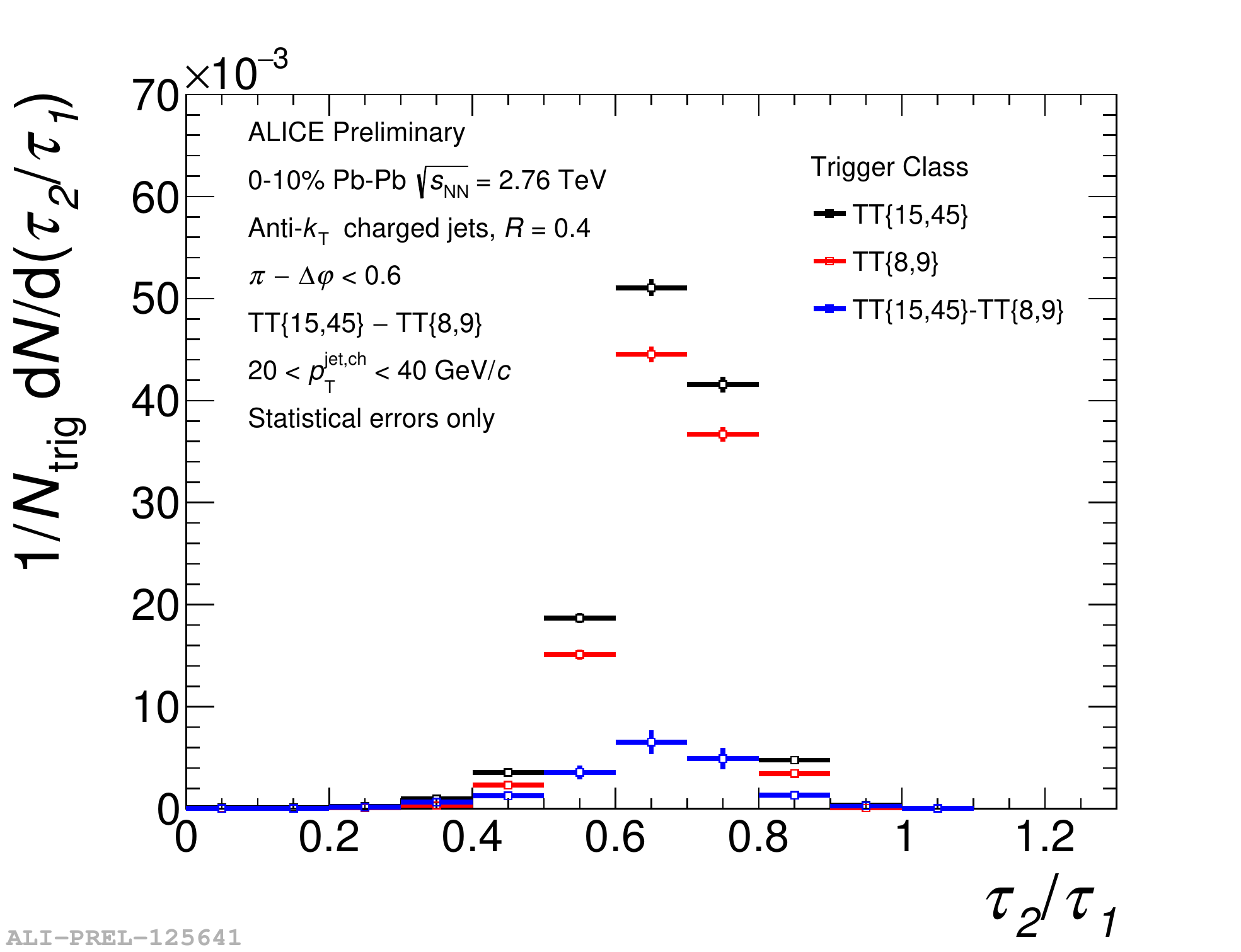}
	\includegraphics[width=.46\textwidth]{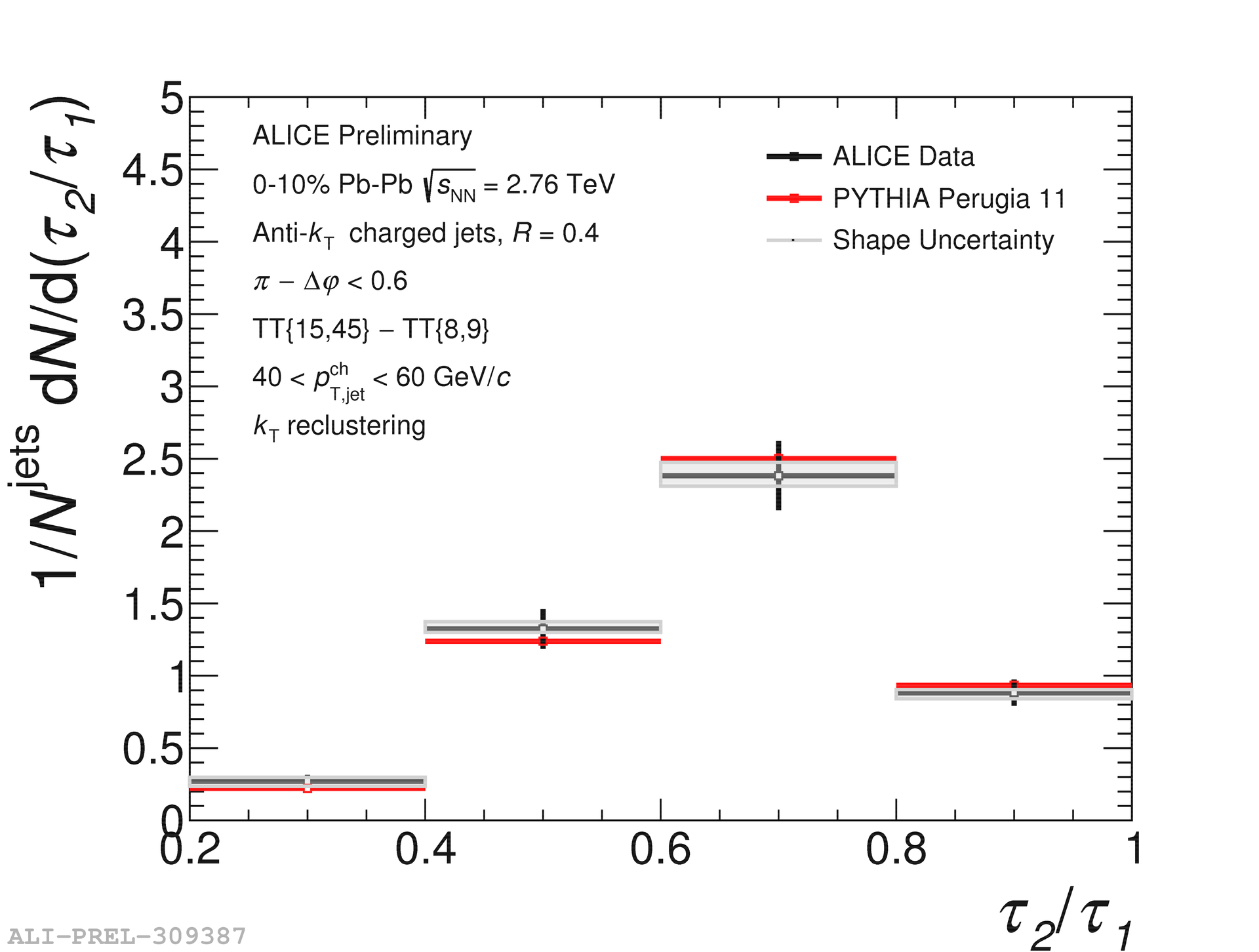}
	\caption[]{Left: The trigger-normalised $\tau_2/\tau_1$ distributions in a signal and reference class, and the difference between the two. Right: The trigger-normalised $\tau_2/\tau_1$ distribution in $\PbPb$ collisions compared to a PYTHIA reference.}  
	\label{fig:subjettiness}
\end{figure}

In order to get a combinatorial background-free distribution of low-$\pt$ jets with low fragmentation bias, a similar technique as described in Section \ref{sec:intro} is used. As shown in figure \ref{fig:subjettiness} (left) a `reference' trigger track recoil jet distribution is subtracted from a `signal' trigger track distribution to obtain the $\tau_2 / \tau_1$ distribution of true jets, free from fragmentation bias. Figure \ref{fig:subjettiness} (right) shows the $\tau_2 / \tau_1$ distribution in $\PbPb$ collisions in comparison with the same distribution in PYTHIA. No modification of two-prongness with respect to PYTHIA is seen within the experimental uncertainties. Different reclustering algorithms are also explored which are sensitive to different properties of the jet splitting, see \cite{subjettiness-these} for more information.

\section{Di-jet azimuthal correlations}
\label{sec:cprrelation}


The interaction of jets with the Quark-Gluon Plasma can be further studied by measuring the azimuthal correlation of dijets, or in this case, the azimuthal correlation between a trigger hadron and a corresponding recoiling jet. This is measured by the angle between the trigger hadron and recoiling jet, denoted $\Delta \varphi$. The motivation for studying this observable is two-fold:

\begin{enumerate}
	\item The broadening of the peak of the away side distribution (at $\Delta \varphi \sim \pi$ ) with respect to vacuum expectation is sensitive to soft multiple scattering in-medium. Since the angular deflection can be related to the change in the momentum transverse to the direction of the initial parton, this could give direct constrains to the transport coefficienct $\hat{q}$ \cite{Chen:2016vem}.
	\item The shape of the tails of the distribution at large angles away from $\pi$ can be used to study the rate of large angle scattering in the QGP. This can arise from resolving the weak degrees of freedom in the Quark-Gluon Plasma, and evidence of large angle scattering could give evidence of a quasiparticle nature of the plasma \cite{moliere-these}.
\end{enumerate}

The background-subtracted hadron+jet azimuthal distribution was measured in $\PbPb$ collisions at ALICE with Run 1 data \cite{Adam:2015doa}. The rate of large angle scattering showed no deviation from the vacuum expectation within the experimental uncertainties, though this measurement was statistically limited. Recent theoretical work (see e.g. \cite{moliere-these,Gyulassy:2018qhr}) has suggested that low-$\pt$ jets are most sensitive to such effects and additional, higher statistics measurements are underway.

\begin{figure}[hbt]
	\centering
	\includegraphics[width=.42\textwidth]{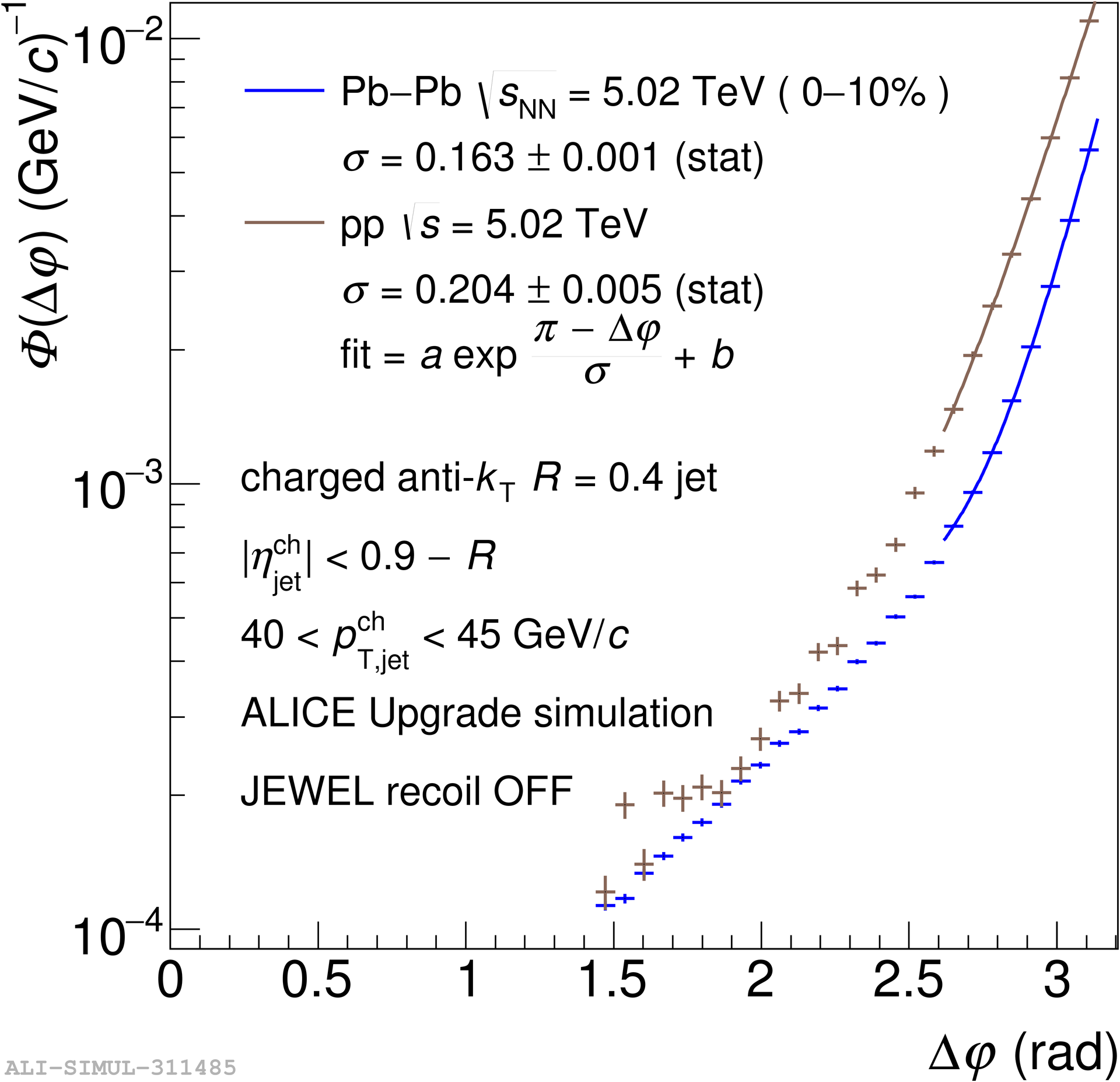}
	\includegraphics[width=.42\textwidth]{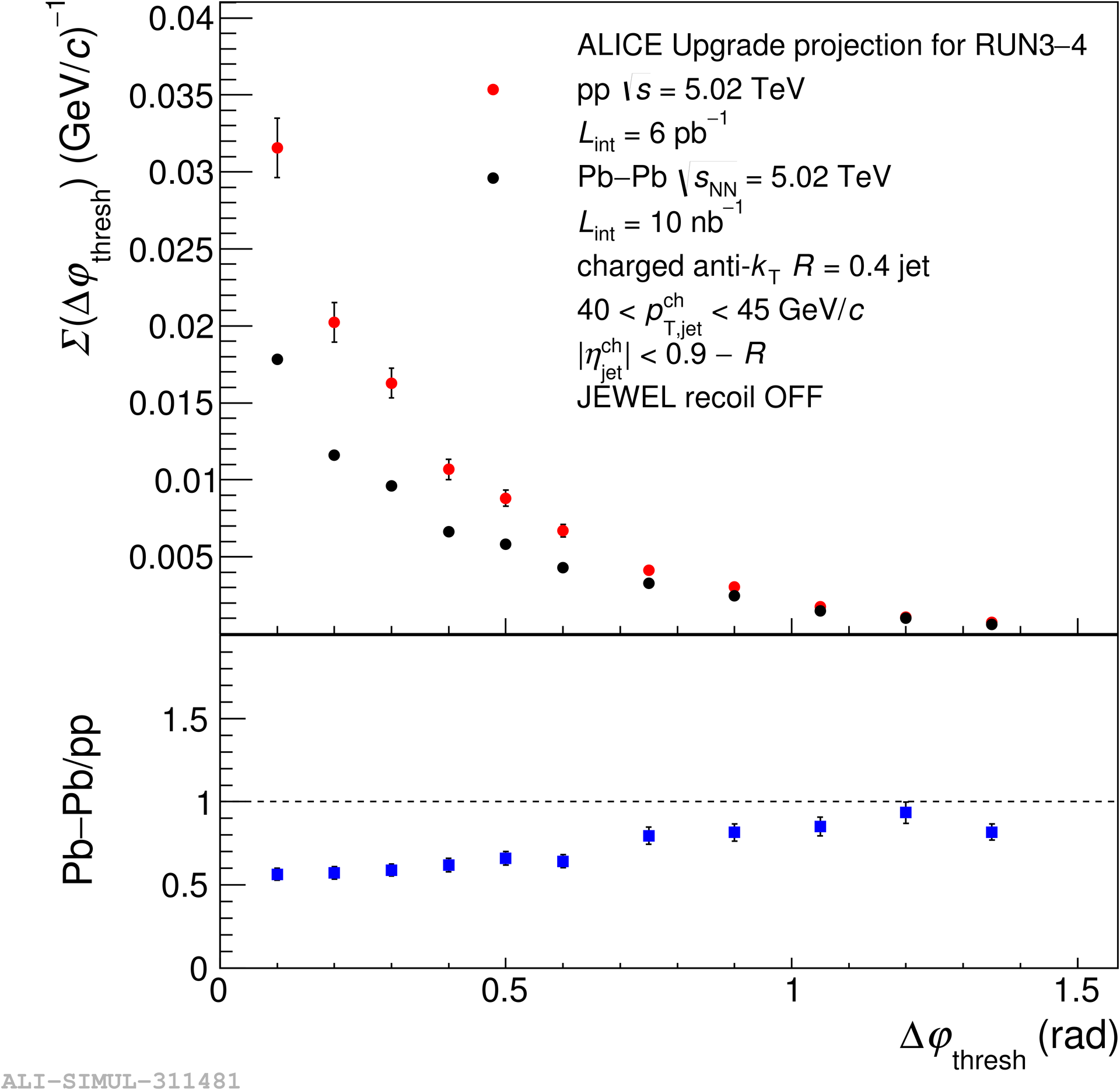}
	\caption[]{Left: Projection of the background-corrected azimuthal hadron-jet distribution in pp and $\PbPb$ collisions in Run 3/4. Right: The integral of this distribution at large angles as a function of the threshold angle of integration $\Delta \varphi_\mathrm{thresh}$, and its ratio in $\PbPb$ and pp collisions.}  
	\label{fig:azimuthal-jets}
\end{figure}

The reach of a hadron+jet measurement in Run 3/4 of the LHC has been assessed. Central $\PbPb$ and pp collisions are simulated with the JEWEL model \cite{Zapp:2013vla}. Figure \ref{fig:azimuthal-jets} (left) shows the background-corrected azimuthal distribution of jets recoiling from a high-$\pt$ trigger hadron, with the expected statistics of Run 3/4. The difference between the large-angle recoil jet yield in pp and $\PbPb$ collisions is then studied by integrating this distribution at large angles away from $\pi$, between $\pi / 2$ and a theshold angle $\Delta \varphi_{\mathrm{threshold}}$, defining $\Sigma(\Delta\varphi_{\mathrm{thresh}}) = \int_{\pi/2}^{\pi-\Delta\varphi_{\mathrm{thresh}}} \mathrm{d}\Delta\varphi [\Phi(\Delta\varphi) ] $.


Figure \ref{fig:azimuthal-jets} (right) shows the $\Sigma(\Delta\varphi_{\mathrm{thresh}})$ distribution in $\PbPb$ and pp collisions in JEWEL, and the ratio between the two systems.
The statistical precision of the ratio at $\Delta \varphi_{\mathrm{threshold}}$ is around 5\% (dominated by the uncertainty of the pp reference), so a deviation in the large-angle yield from vacuum will be able to be resolved to approximately this accuracy. Theoretical calculations predict deviations of similar magnitude \cite{Gyulassy:2018qhr}, so a measurement in Run 3/4 promises to resolve different pictures of the micro-structure of the medium.

\bibliographystyle{JHEP}
\bibliography{my-bib-database}


\end{document}